\documentclass[conference]{IEEEtran}
\IEEEoverridecommandlockouts

\usepackage{tikz}
\usepackage{amsmath}
\usepackage{amsthm}
\usepackage{amssymb}
\usepackage[]{graphicx}
\usepackage{subfig}
\usepackage{algorithm}
\usepackage[noend]{algpseudocode}

\usepackage{booktabs}
\usepackage{bm}
\usepackage{multirow} 
\usepackage{multicol}
\usepackage{flushend}
\usepackage{makecell}
\usepackage{threeparttable}
\usepackage{diagbox}
\usepackage{footnote}
\usepackage{lipsum}
\usepackage{xcolor}
\usepackage{enumitem}
\usepackage{fancyhdr}
\usepackage{listings}
\definecolor{difftitle}{HTML}{000099}
\definecolor{diffstart}{HTML}{660099}
\definecolor{diffincl}{HTML}{006600}
\definecolor{diffrem}{HTML}{AA3300}
\lstdefinelanguage{diff}{
    backgroundcolor=\color{white},  
    basicstyle=\ttfamily\small,
    morecomment=[l][\color{difftitle}]{diff},
    morecomment=[l][\color{diffstart}]{@@},
    morecomment=[f][\color{diffincl}]{+},
    morecomment=[f][\color{diffrem}]{-},
    columns=fullflexible,
    tabsize=4,
    breaklines=true,
    captionpos=b, 
    frame=none,
}    

\usepackage{hyperref}
\hypersetup{
    colorlinks=true,
    linkcolor=blue,
    filecolor=magenta,      
    urlcolor=blue,
    }
\usepackage{cite}
\usepackage{amsmath,amssymb,amsfonts}
\usepackage{algorithm}
\usepackage{algpseudocode}
\usepackage{graphicx}
\usepackage{textcomp}
\usepackage{xcolor}
\usepackage{colortbl}
\usepackage{makecell}
\def\BibTeX{{\rm B\kern-.05em{\sc i\kern-.025em b}\kern-.08em
    T\kern-.1667em\lower.7ex\hbox{E}\kern-.125emX}}

\lstdefinestyle{lst}{
    float=tp,
    floatplacement=tbp,
    numbers=left, 
    numberstyle=\scriptsize, 
    numbersep = 5pt,
    framexleftmargin = 0in,
    framexrightmargin = 0in,
    xleftmargin = 0.18in,
    xrightmargin = 0.1in,
    basicstyle=\ttfamily\scriptsize, 
    frame=lines,
    showtabs=true,
    showspaces=true,
    showstringspaces=false,
    literate={\ }{{\ }}1,
}

\def\footnoterule{\relax%
  \kern-0pt
  \hbox to \columnwidth{\hfill\vrule width \columnwidth height 0.5pt\hfill}
  \kern3pt}
\makeatother

\begin{document}

\newcommand{\db}{PySecDB}
\newcommand{\dbone}{base}
\newcommand{\dbtwo}{pilot}
\newcommand{\dbthree}{augmented}
\newcommand{\gnn}{SCOPY}
\newcommand{\cpg}{CommitCPG}
\newcommand{\emb}{CommitEmbedding}

\title{Exploring Security Commits in Python}


\author{Shiyu Sun$^{*}$,
Shu Wang$^{*}$,
Xinda Wang$^{*}$,
Yunlong Xing$^{*}$,
Elisa Zhang$^{\dagger}$,
Kun Sun$^{*}$\\
$^{*}$George Mason University,
$^{\dagger}$Dougherty Valley High School\\
\{ssun20, swang47, xwang44, yxing4, ksun3\}@gmu.edu, elisaz.ca@gmail.com}

\maketitle
\thispagestyle{fancy}
\pagestyle{fancy}

\begin{abstract}

Python has become the most popular programming language as it is friendly to work with for beginners. However, a recent study has found that most security issues in Python have not been indexed by CVE and may only be fixed by ``silent" security commits, which pose a threat to software security and hinder the security fixes to downstream software. It is critical to identify the hidden security commits; however, the existing datasets and methods are insufficient for security commit detection in Python, due to the limited data variety, non-comprehensive code semantics, and uninterpretable learned features.
In this paper, we construct the first security commit dataset in Python, namely \db{}, which consists of three subsets including a base dataset, a pilot dataset, and an augmented dataset. The base dataset contains the security commits associated with CVE records provided by MITRE.
To increase the variety of security commits, we build the pilot dataset from GitHub by filtering keywords within the commit messages.
Since not all commits provide commit messages, we further construct the augmented dataset by understanding the semantics of code changes.
To build the augmented dataset, we propose a new graph representation named \cpg{} and a multi-attributed graph learning model named \gnn{} to identify the security commit candidates through both sequential and structural code semantics. The evaluation shows our proposed algorithms can improve the data collection efficiency by up to 40 percentage points.
After manual verification by three security experts, \db{} consists of 1,258 security commits and 2,791 non-security commits.
Furthermore, we conduct an extensive case study on \db{} and discover four common security fix patterns that cover over 85\% of security commits in Python, providing insight into secure software maintenance, vulnerability detection, and automated program repair.

\end{abstract}



\section{Introduction}


According to the TIOBE index~\cite{tiobe2}, Python overtakes Java and C as the most popular programming language as of  April 2023. 
However, a recent study~\cite{ruohonen2021large} reveals that among 749K security issues of 197K Python packages in PyPI~\cite{pypi}, only 1,232 vulnerabilities are reported to CVE~\cite{cve} and only 556 of them have public fixes.
Thus, most security issues are not indexed and may only be resolved by ``silent" fixes, without explicit log messages indicating the vulnerabilities. 

The hidden security fixes pose a threat to the security and privacy of users, since attackers may exploit the undisclosed vulnerabilities to comprise the unpatched software systems.
Particularly, Python is friendly to beginners; however, with limited security knowledge, learners may be unable to determine if an upstream commit is intended to address a vulnerability and hence neglect a critical security fix.
The implicit security commits can impair software maintenance and evolution since the downstream developers may not be aware of the criticality of these security commits.
Therefore, it is vital to identify the hidden security commits among all Python commits.


The existing datasets and methods are insufficient for security commit detection in Python. To identify security commits, researchers first build commit datasets~\cite{wang2021patchdb,ponta2019manually} and then either extract features manually from commit messages/code changes~\cite{li2017large,wang2021patchdb,kim2017vuddy} or learn features automatically via deep learning models, e.g., recurrent neural networks (RNN)~\cite{wang2021patchrnn,li2016vulpecker}, Transformers~\cite{zhoucolefunda,zhou2021finding}, and graph neural networks (GNN)~\cite{wang2022graphspd,wu2022enhancing}.
However, these solutions have three main constraints, namely, limited data variety, non-comprehensive code semantics, and uninterpretable learned features.
First, existing works~\cite{zhou2021spi,wang2021patchdb,ponta2019manually} only construct security commit datasets in C/C++ or Java, and there is no available Python dataset for security commit research.
Second, the existing feature extraction methods cannot be directly applied to Python security commit detection. The manually extracted features~\cite{li2017large,wang2021patchdb} are language-dependent and cannot be directly migrated to the Python language.
Also, the existing deep learning features~\cite{wang2021patchrnn,zhou2021finding,zhou2021spi,wang2022graphspd} are incapable of integrating both the sequential and structural semantics of code since the RNN and Transformer-based models simply treat code as a natural language~\cite{hochreiter1997long, feng2020codebert} and the GNN-based models only focus on the code dependencies~\cite{wang2022graphspd, wu2022enhancing}.
Third, though manually extracted features are interpretable, feature extraction methods  can only provide limited accuracy. In contrast, deep learning based methods may yield better accuracy, but their learned features are uninterpretable.



To tackle the above challenges, we construct the first security commit dataset in Python, named \db{},\footnote{This dataset is released in \href{https://github.com/SunLab-GMU/PySecDB}{https://github.com/SunLab-GMU/PySecDB}.} by collecting the samples from CVE records and filtering the commits from GitHub. It consists of three subsets: a base dataset, a pilot dataset, and an augmented dataset.
%
We first build a base dataset by collecting the security commits associated with CVE IDs~\cite{cve}. Since the CVE records on Python programs are limited, we observe that  only 46\% of them provide the corresponding security commits and more security commits fall in the wild silently, without being indexed by CVE.
%
%
To enrich security commits for covering more vulnerability types, we construct a pilot dataset by filtering GitHub commits. Since only 6-10\% of GitHub commits are related to security fixes~\cite{wang2021patchdb}, filtering commit messages using relevant keywords can efficiently narrow down the list of security commit candidates.
The security keywords are automatically extracted from the commits in the base dataset by the Latent Dirichlet Allocation (LDA) method~\cite{blei2003latent}. Then, three  security experts manually verify the candidate commits and build the pilot dataset.
Well-documented commit messages are required for the pilot dataset construction, but not all commits provide commit messages.

To include more diverse security commits that do not provide sufficient commit messages, we extend the base and pilot datasets with an augmented dataset by considering the semantics of code revisions.
We develop a new commit graph representation named \cpg{} and a graph learning model named \gnn{} 
to capture the semantics of code changes. Inspired by~\cite{y2014cpg}, our \cpg{} is constructed by merging the code property graphs of the previous and current versions.
In \cpg{}, each node presents a statement with its version information; each edge preserves the semantic level dependency between two statements.
To reduce analysis overhead, we perform program slicing~\cite{weiser1984program} over \cpg{}s to remove the nodes/edges irrelevant to the commits.
Given a \cpg{}, \gnn{} embeds the node statements using CodeBERT~\cite{feng2020codebert} and embeds the edge attributes as one-hot vectors to contain the edge versions and the syntax/dependency relationships.
Then, a graph convolutional network with multi-head attention is trained over the base and pilot datasets. Finally, we apply \gnn{} to identify the security commit candidates from the wild and build the augmented dataset after manual verification.



To enhance the variety of commits beyond the base dataset, we construct the pilot and augmented datasets over popular repositories.  We evaluate the efficiency of data collection using the ratio of security commits to the total number of candidates. Compared with random selection, the keyword filtering method and \gnn{} can improve the efficiency by 30 and 40 percentage points when constructing the pilot and augmented datasets, respectively.
In total, \db{} contains 1,258 security commits associated with 119 distinctive CWEs across 351 repositories, providing sufficient diversity in vulnerability types and application scenarios.
We also find that unique patterns exist in the pilot and augmented datasets, respectively, since these two datasets are built from different perspectives, i.e., commit messages and code changes.


To facilitate software maintenance, we conduct an extensive case study on the security commits in \db{} and discover four common security fix patterns, i.e., {\em add or update sanity checks}, {\em update API usage}, {\em update regular expressions}, and {\em restrict security properties}. 
First, security commits often include sanity checks (i.e., verify if certain conditions are true) to secure critical operations, especially in the authentication and authorization scenarios. 
Second, since Python provides multiple pre-built packages, security commits can address vulnerabilities by replacing APIs, e.g., APIs related to strings, paths, and commands.
Third, security commits can handle secure escapes by updating regular expressions, which protect software from being injected by shell commands, SQL queries, and web scripts.
Fourth, security commits can update security properties (e.g., security flags, restriction arguments, and security decorators) to ensure the effectiveness of security mechanisms or policies.
These fix patterns can be generalized and formulated into intermediate representations to facilitate secure software development and automated program repair.

In summary, our paper makes the following contributions:
\begin{itemize}
    \item We construct the first security commit dataset in Python by screening CVE records and GitHub commits. 
    \item We design a keyword filtering method to identify the potential security commits based on the commit messages.
    \item Based on code changes, we propose a new commit graph representation \cpg{} and a graph learning model \gnn{} to locate the security commit candidates.
    \item To facilitate software maintenance, we discover four common security fix patterns, which  provide insights in vulnerability detection and program repair in Python.    
\end{itemize}


\section{Background and Related Work}
\label{sec:background}\label{sec:relatedwork}
\vspace{-0.05in}

\lstdefinestyle{lst}{
    float=tp,
    floatplacement=tbp,
    numbers=left, 
    numberstyle=\scriptsize, 
    numbersep = 5pt,
    framexleftmargin = 0in,
    framexrightmargin = 0in,
    breaklines = true,
    xleftmargin = 0.18in,
    xrightmargin = 0.1in,
    basicstyle=\ttfamily\scriptsize, 
    frame=lines,
    showtabs=true,
    showspaces=true,
    showstringspaces=false,
    literate={\ }{{\ }}1,
    aboveskip=+0.15in,
    belowskip=-0.00in,
}

\begin{lstlisting}[
language=diff, 
style=lst,
caption=An example of security commit (CVE-2021-27213).,
label={lst:security commit},
mathescape=true
]
 $\textbf{commit dbeb87afefdb63de2f4cff69b6f10c5965d14b54}$ 
 Subject: [PATCH] Fixed code execution bug using SafeLoader()
 diff --git a/pystemon/config.py b/pystemon/config.py
 @@ -315,7 +315,7 @@ def _load_yamlconfig(self
 yamlconfig = None
 ...
 for includes in yamlconfig.get("includes", []):
    try:
        logger.debug("... '{0}'".format(includes))
-       yamlconfig.update(yaml.load(open(includes)))
+       yamlconfig.update(yaml.safe_load(open(includes)))
    except Exception as e:
        raise PystemonConfigException("failed to load '{0}': {1}".format(includes, e))
 return yamlconfig

\end{lstlisting}

\lstdefinestyle{lst}{
    float=tp,
    floatplacement=tbp,
    numbers=left, 
    numberstyle=\scriptsize, 
    numbersep = 5pt,
    framexleftmargin = 0in,
    framexrightmargin = 0in,
    breaklines = true,
    xleftmargin = 0.18in,
    xrightmargin = 0.1in,
    basicstyle=\ttfamily\scriptsize, 
    frame=lines,
    showtabs=true,
    showspaces=true,
    showstringspaces=false,
    literate={\ }{{\ }}1,
    aboveskip=-0.00in,
    belowskip=-0.2in,
}

\begin{lstlisting}[
language=diff, 
style=lst,
caption=An example of non-security commit.,
label={lst:non-security commit},
mathescape=true
]
 $\textbf{commit 4cd1067faf3df14dbbe7eb6de2bd7693d5cd829a}$ 
 diff --git a/IPython/lib/security.py b/IPython/lib/security.py
 @@ -109,7 +109,7 @@ def passwd_check(hashed_passphrase, passphrase):
     except ValueError:
         $\textbf{return}$ False
-    if len(pw_digest) == 0 or len(salt) != salt_len:
+    if len(pw_digest) == 0:
         $\textbf{return}$ False
\end{lstlisting}

\subsection{Security and Non-security Commits} 
On the version control platforms such as GitHub, a commit is 
mainly composed of two parts: a set of code changes between two versions and a descriptive message including the subject line and body (if any). In Listing~\ref{lst:security commit}, Lines 5-14 are the source code changes and Line 2 presents a commit message with only one subject line.

A security commit includes code changes made to a software codebase, addressing a security vulnerability defined by an individual Common Weakness Enumeration (CWE) Specification~\cite{CWE_slice}. 
Security commits are typically critical updates that need to be applied as soon as possible to prevent attackers from exploiting vulnerabilities. 
List~\ref{lst:security commit} shows a security commit example fixing the vulnerability CVE-2021-27213 by replacing \texttt{\small yaml.load()} with \texttt{\small yaml.safe\_load()} to load the content in a safer way.
Non-security commits are the changes made to the software codebase that do not relate to security issues. 
These changes include fixing non-security-related bugs, adding new features, improving performance, and updating documentation. 
Typically, non-security commits are not as urgent as security commits and can be applied later without affecting software security. 
In List~\ref{lst:non-security commit}, a non-security commit removes the unnecessary check on password salt length.

\subsection{Security Commit Datasets} 
Security commits provide plentiful information on both the existing vulnerabilities and the corresponding fixes. 
Thus, researchers construct such datasets for security commit detection and automated program repair~\cite{ponta2019manually, fan2020ac, wang2021patchdb, nikitopoulos2021crossvul, bhandari2021cvefixes, chen2023diversevul}.
However, existing datasets only focus on specific projects \cite{ponta2019manually, zhou2021spi} or contain limited security commits associated with CVEs~\cite{fan2020ac, bhandari2021cvefixes}.
Although some researchers also consider both commits indexed by NVD and silent fixes~\cite{wang2021patchdb, nikitopoulos2021crossvul, chen2023diversevul}, 
they solely investigate the commits in C/C++ and Java, neglecting the popularity of Python. 
Besides, 
the existing works adopt language-dependent security-related features, which cannot be directly migrated for collecting security commits in Python. 

\subsection{Security Commit Detection} Numerous commits are submitted to GitHub every day, while 6-10\% of them are silent security fixes~\cite{li2017large, wang2019detecting}. To identify security commits automatically and effectively, \cite{zhou2017automated} analyzes the natural language description of commit messages and bug reports. However, this approach only relies on well-maintained documentation, which is impractical for detecting silent security patches. 
Thus, some researchers detect security commits by extracting code features manually~\cite{wang2021patchdb, sawadogo2020learning}.
Wang et al.~\cite{wang2021patchrnn} ensemble two BiLSTM models to learn not only the commit message but also the code changes. 
Similarly, SPI~\cite{zhou2021spi} adopts LSTM to learn the representation of commit message and utilizes CNN to learn the representation of code revision. 
As the prevalence of applying large language model on code analysis, CodeBERT~\cite{feng2020codebert} is fine-tuned to learn the semantics of code changes~\cite{zhou2021finding}. 
Zhou et al.~\cite{zhoucolefunda} increase the fix data at the function level and then generalize the code change semantic with contrasting learning.
 


To preserve the inherent structural semantics of code, 
CLozoya et al.~\cite{cabrera2021commit2vec} and Wu et al.~\cite{wu2022enhancing} employ BiLSTM to learn the representation of commits from their AST paths. Wang et al.~\cite{wang2022graphspd} propose GraphSPD to represent C/C++ commits with code property graphs and learn the representation with GNN.



\subsection{Novelty of Our Study}

\noindent{\bf Dataset.} We build the first security commit dataset in Python, which contains 1,258 security commits and 2,791 non-security commits extracted from over 351 popular GitHub projects, covering 119 more CWEs.
Different from the SPI~\cite{zhou2021spi} based on keyword filtering and PatchDB~\cite{wang2021patchdb} based on code similarity, we consider both commit message and code changes so that the dataset can cover more diverse security commits, especially for the commits without any clear commit message.


\noindent{\bf Commit Representation.} 
To better represent the commits, we preserve the code structure via the graph representation \cpg{} and also consider the sequential information via CodeBERT~\cite{feng2020codebert}.
However, the existing works either consider the sequential information~\cite{zhou2017automated, wang2021patchrnn, zhou2021finding, zhoucolefunda} or keep the structural semantics~\cite{wang2022graphspd}.




\noindent{\bf Commit Understanding.} 
We conduct a comprehensive study on understanding how the security commits fix Python vulnerabilities. 
The commit patterns can be used to generalize vulnerability fix schemes, which may enhance software maintenance and provide insight into automated program repair.
\section{Data Collection}
\label{sec:design}

\begin{figure}[t]
    \centering
    \includegraphics[width=3.33in]{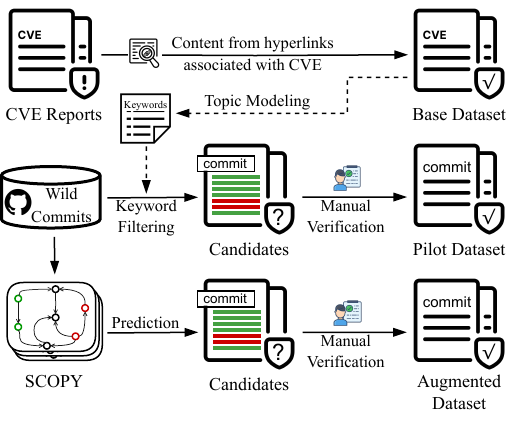}
    \vspace{-0.05in}
    \caption{Overview of collecting three datasets.}
    \vspace{-0.15in}
    \label{fig:system}
\end{figure}

Our dataset consists of three sections: \textit{a) base dataset, b) pilot dataset}, and \textit{c) augmented dataset}. Figure~\ref{fig:system} illustrates the composition and construction procedure of~\db{}. 
We first form the base dataset by collecting the commits associated with CVE records indexed by MITRE. Yet, less than 50\% of CVE records have published their security commits. 
To introduce more code semantics, we further build the pilot dataset by filtering the wild GitHub commits that have pre-defined security keywords in their commit messages. 
However, not all commits contain well-maintained commit messages that precisely describe the rationales of changed code.
Thus, we consider directly mining the most critical part of commits, i.e., the source code changes. 
To the end, we propose an intermediate commit representation (i.e., \cpg{}) and design a dependency-aggregation graph neural network (i.e., \gnn{}) to capture the inherent sequential and structural semantics of code changes.
Trained with the base and pilot datasets, \gnn{} is able to further build the augmented dataset by pinpointing the silent security commits from the wild.

\subsection{Base Dataset Collection}
We build the base dataset 
according to the CVE records~\cite{cve}.
The first step is to retrieve the vulnerabilities that have already been indexed with CVE IDs. Then, we parse the vulnerability reports and crawl the corresponding commits via the provided reference hyperlinks. It comes to our attention that the collected commits may contain some noise, e.g., changelog, test case, refactoring, and renaming. 
After excluding these unrelated documents, we obtain 729 security commits to form the base dataset; meanwhile, we collect the excluded commits as the non-security subset in the base dataset and expand it by manually identifying the commits that add new features or perform refactoring, linting, and version updates.

\subsection{Pilot Dataset Collection}
\label{db:pilot}


After examining all the indexed CVE records (as of 01/27/2023), only 46\% of them contain the corresponding security fixes. Therefore, the limited samples in the base dataset may not provide adequate syntactic and semantic information.
That means, only with the base dataset, we are unable to train a robust model for capturing a wide variety of security commits in the real world.
Given the fact that a majority number of security patches are silently committed without reporting to the MITRE~\cite{wang2019detecting,zhou2021spi}, we propose to enrich the security commits with the pilot dataset collected from GitHub, i.e., the most common OSS hosting platform. 

The pilot dataset is constructed by keyword filtering with humans in the loop. 
The list of security-related keywords is built automatically by analyzing the CVE descriptions, CWE types (if exist), and commit messages.
We determine the final keywords by calculating the word frequency and evaluating the correlation between the keywords and security commits. 
To obtain the security subset of the pilot dataset, we locate and manually verify the security commit candidates that contain the pre-defined security keywords in the commit messages; the excluded commits are collected as the non-security subset.


\noindent{\bf Security Keyword Extraction. }
For each security commit collected from the CVE records, we generate its security impact summary by combining the commit message, the CWE information (if exists), and the CVE report.
After generating the summary, we conduct 1-gram, 2-gram, and 3-gram tokenization. 
Then, we consider the frequency of each token and the correlation of each token with security and non-security commits.
We set the frequency threshold and derive the list of security-related keywords, as shown in Table~\ref{tab:keywords}. 
Then, we determine the security commit candidates by checking if the wild GitHub commits contain any security keywords in their commit messages.
These candidates will be manually verified.


\begin{table}[t]
\begin{center}
\caption{Security-related keywords for commit filtering.}
\label{tab:keywords}
\resizebox{\linewidth}{!}{
\begin{tabular}{c|c}
\toprule
{\bf \#Tokens} & \textbf{Keywords} \\ 
\midrule
{1-gram} & {\begin{tabular}[c]{@{}c@{}}
attack, bypass, CVE, DoS, exploit, injection, \\ leakage, malicious, overflow, smuggling, \\ spoofing, unauthorized, underflow, vulnerability
\end{tabular}} \\ 
\midrule
{2-gram} & {access control, open redirect, race condition} \\ 
\midrule
{3-gram} & {denial of service, out of bound,  dot dot slash} \\ 
\bottomrule
\end{tabular}
}
\end{center}
\vspace{-0.2in}
\end{table}

\noindent{\bf Manual Verification.} 
We hire three security experts to manually verify the security commit candidates. 
To guarantee the data quality and minimize false positives, the experts are required to follow our two-step labeling procedure strictly. 
First, each expert tags the commits independently with the labels: security, non-security, or unsure. 
Then, they gather together to discuss each disagreement and reach a consensus on each uncertain candidate. 
Only the security commits with 100\% agreement will be included in the pilot dataset. In total, it takes 48 man-hours to finish the labeling work.

The proposed keyword filtering mechanism reduces the workload and time for manual verification; meanwhile, the human-in-the-loop ensures the quality of the pilot dataset. More details on labeling efficiency are shown in Section~\ref{results:efficiency}.

\subsection{Augmented Dataset Construction}

The pilot dataset overlooks the commits that lack security keywords in the commit messages, while these commits may provide additional variants in syntax and semantics. 
Therefore, we propose to further augment our dataset. 
While the pilot dataset is collected based on commit messages, we build the augmented dataset by only analyzing the source code changes.
Different from existing works that simply regard source code as sequential data~\cite{wang2021patchrnn,zhou2021spi}, we present a commit graph representation named CommitCPG and a graph learning-based model \gnn{} to capture the inherent structural information.

\vspace{0.05in}
\noindent \textit{1) CommitCPG: Graph Representation for Commits}
\vspace{0.02in}

\begin{figure*}[h]
    \centering
    \includegraphics[width=0.85\linewidth]{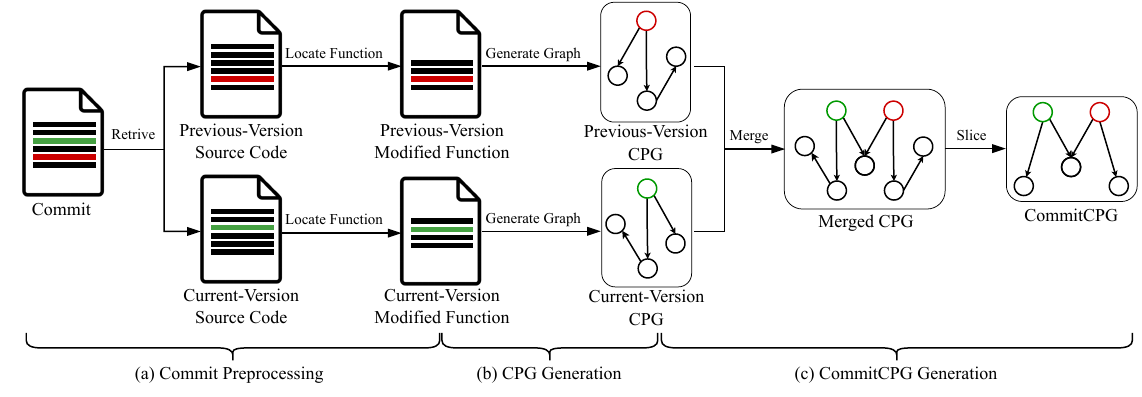}
    \vspace{-0.05in}
    \caption{The generation process of \cpg{}s from commits.}
    \label{fig:commitCPG}
    \vspace{-0.2in}
\end{figure*}

To preserve the inherent structure of source code and the modified content between two versions, we propose a graph-based commit representation called \cpg{}, which offers essential syntactic and semantic information for comprehensive commit understanding. 
Code property graph (CPG)~\cite{y2014cpg} is a program representation that contains abstract syntax trees (AST), control-flow graphs (CFG), and program dependence graphs (PDG), providing a more comprehensive view for code static analysis, compared with traditional sequential structure adopted by NLP-based works~\cite{guo2020graphcodebert}.
In Figure \ref{fig:commitCPG}, we first preprocess the raw commits by excluding the irrelevant functions. Inspired by \cite{wang2022graphspd}, we then merge the CPGs \cite{y2014cpg} constructed from the previous and current versions by aligning the unchanged statements. Next, we adopt a code slicing method to retain the crucial context-related code snippets, which are not changed directly by commits but can assist us to understand the reason and the effects of  code changes.

\noindent{\bf Commit Preprocessing.} 
To generate the CPG for each code version, we need to retrieve the source code of the previous version and the current version, respectively.
To reduce the overhead of CPG generation, we only focus on the modified files instead of the whole project. 
Then, we extract the functions with code revisions as well as the modified global statements. 
To achieve this goal, Joern parser \cite{y2014cpg} is applied to detect all relevant functions and their corresponding scopes. 
We only retain the functions whose scope overlaps with the modified lines in the commits.
For example, in List~\ref{lst:security commit}, we will only keep the content of function \texttt{\small \_load\_yamlconfig()}.

\noindent{\bf CPG Generation for Previous and Current Versions.} 
With the extracted source code of two versions, we employ Joern~\cite{y2014cpg} to generate the CPGs for both versions. 
A CPG can be described as $(V, E)$, where $V$ is the node set and $E$ is the edge set.
$V$ is comprised of multiple 5-tuples $(id, func\_name, file\_name, version, code)$, which contain the information of each node.  
The node version, represented by $version \in \{previous, current\}$, reflects if the code line belongs to the previous version or current version.
The directed edge set $E$ is made up of 4-tuples $(id_1, id_2, type, version)$, where $id_1$ and $id_2$ denote the start and end node IDs, respectively.
The edge type, represented by $type \in \{AST, CDG, DDG\}$, specifies if the edge belongs to the AST or control/data dependency graphs. 
The edge $version$ is consistent with the node's version.

\noindent{\bf CPG Merging and CommitCPG Slicing.}
We first generate a unified commit graph by fusing the CPGs of two versions according to each node pair.
Then, we update the representation of the merged CPG by two sets: $(V', E')$. 
The node set $V '$ is comprised of 5-tuples, which are denoted as $(id, func\_name, file\_name, version, code)$. 
$id$ is updated so that each node has a unique identifier and the node version is changed to $version \in \{current, previous, unchanged\}$, representing if the code in this node belongs to the current, previous, or both commits. 
The directed edge set $E'$ is made up of 4-tuples $(id_1, id_2, type, version)$, where $id_1$, $id_2$, and $type$ stay the same as $E$. The edge $version$ will be updated as $unchanged$ if both connected nodes are unchanged nodes.

To reduce the noise introduced by irrelevant nodes and emphasize the semantics of code changes, we generate the \cpg{} by a bi-directional program slicing \cite{weiser1984program}, i.e., forward and backward slicing. 
Backward slicing is to reason the code changes, while forward slicing is to locate the statements affected by the commit. 
For example, in List~\ref{lst:security commit}, if we set the deleted statement (Line 10) as a backward slicing criterion, the slicing results include Lines 5, 7, and 8; if we set the added statement (Line 11) as a forward slicing criterion, the slicing results contain Line 14.
After we conduct code slicing over control/data dependency, we can obtain \cpg{} by only retaining all the nodes of modified and sliced statements (i.e., Line 7, 8, 10, 11, and 14) along with the traced edges.


\vspace{0.05in}
\noindent \textit{2) \gnn{}: Graph Learning for Commits}
\vspace{0.02in}

\begin{figure*}[h]
    \centering
    \includegraphics[width=0.9\linewidth]{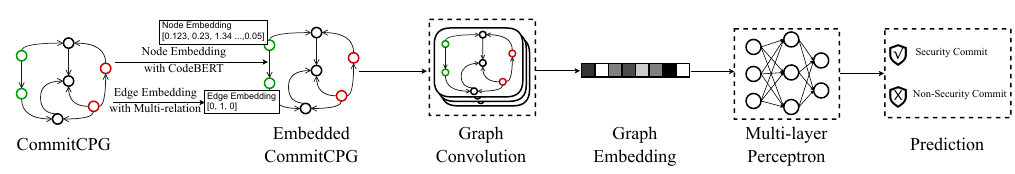}
    \vspace{-0.05in}
    \caption{The workflow of \gnn{} that identifies security commits via \cpg{}s.}
    \vspace{-0.2in}
    \label{fig:model}
\end{figure*}

Figure~\ref{fig:model} illustrates the workflow of \gnn{}, our proposed graph-based network to identify security commits in Python. \gnn{} contains two steps: (i) node embedding with CodeBERT and edge embedding with dependency-aggregation mechanism, (ii) graph convolution with multi-head attention.

\noindent{\bf \cpg{} Embedding.} To feed the \cpg{} to our \gnn{}, we encode the node and edge attributes into numeric vectors. 
Each node represents a code statement; hence, the node embedding should capture the semantics within the statement.
Thus, we first utilize CodeBERT~\cite{feng2020codebert} to generate token embeddings and grasp the sequential-based semantics.
Then, we obtain the node embedding by aggregating the token embeddings.
In addition, each edge presents the dependency between two nodes; thus edge embedding preserves crucial structural and attribute information.
We generate edge embeddings with 5-dimensional one-hot vectors. 
The first two dimensions are used to embed the structural information and present which code version the edge belongs to.
The last three dimensions are used to embed the attribute information, which indicates if the edge presents control dependency, data dependency, or syntax relationship.






\noindent{\bf Graph Convolution with Multi-Head Attention.} 
After embedding the \cpg{} with a sequential model, we adopt a graph convolutional network with multi-head attention to learn the structural representation of commits. 
To avoid over-smoothing, the number of convolutional layers is limited to 3. 
We feed the embedded \cpg{} into 3 multi-attributed graph convolutional layers. 
The node embeddings of \cpg{} are updated with the neighborhood information from different subgraphs. 
Then, the graph embeddings, i.e., a unified vector representation transformed from all the nodes, edges, and features, can be obtained through graph pooling and vector concatenation.
The graph embeddings learned by the \gnn{} are finally fed into a multi-layer perceptron to determine the likelihood that a commit fixes a security vulnerability, i.e., whether the given commit is a security commit.

To demonstrate the generalization ability of \gnn{}, we utilize the trained \gnn{} to discover more security commits in the wild. 
We directly send the \cpg{} of wild commits into \gnn{}. For the commits labeled as security commits, we adopt a similar process (as described in~\ref{db:pilot}) to manually verify if they are real security commits. The excluded commits composite the non-security subset of the augmented dataset.

\section{Implementation}
\label{sec:implementation}


\subsection{Base Dataset Construction}

The base dataset consists of the commits linked with CVE records, which have been indexed by MITRE~\cite{cve}. MITRE provides hyperlinks of vulnerability fixes for 46\% of CVE entries. We focus on the hyperlinks from GitHub, where each commit is identified with a unique hash value and the hyperlink is in the form: \emph{https://github.com/\{owner\}/\{repo\}/commit/\{hash\}}. {We build the base dataset by downloading the vulnerability fixing commits and removing the commits that are not written in Python or only focus on security-unrelated modifications (e.g., renaming and refactoring).}

\subsection{Pilot Dataset Construction}
We adopt Latent Dirichlet Allocation (LDA)~\cite{blei2003latent}, a topic modeling method, to extract the essential security-related tokens from the commit messages.
Then, a keyword filtering algorithm is applied to exclude non-security commits. 
We download popular open-source repositories in Python and retrieve their commit histories till Jan 27, 2023. For each commit, if it contains at least one proposed keyword, we regard it as a security commit candidate. Later, we manually check these security commit candidates and finalize the dataset.

To facilitate the verification process, we use {PyQt}~\cite{pyqt} to develop a graphical user interface (GUI) that 
visualizes the code changes of each individual commit and stores the verified security commits according to verification results.

\subsection{Augmented Dataset Construction}


To build \gnn{} for further dataset augmentation, we first generate the corresponding \cpg{} for each commit in the base and pilot datasets. Specifically,
we adopt {Joern}~\cite{y2014cpg} to generate CPGs for the code versions before and after applying the commit, respectively. Then, we parse the generated graph files and merge the graphs to build \cpg{}. To achieve the program slicing, we develop a Python script to analyze the control/data dependency and AST information and output a sliced CommitCPG ready to be embedded.

To prepare an embedded graph for \gnn{}, we embed the nodes and edges respectively. We fine-tune CodeBERT~\cite{feng2020codebert} to generate the node embedding dedicated to Python. For the edge embeddings, we apply the one-hot encoding to represent the attributes on each edge. We build the \gnn{} on the deep learning library PyTorch 1.6, which is optimized for tensor computing. We develop and optimize our graph model based on the PyTorch-geometric 1.6 library, which supports deep learning on graphs and other structured data.
Finally, a multiple-layer perceptron (MLP) is used as a binary predictor, which converts the graph embeddings into predicted labels. 
We train \cpg{} with the base and pilot datasets. 
Then, we feed the wild unlabeled commits into the trained \gnn{} and apply manual verification to generate our augmented dataset.

\section{Analysis Results}
\label{sec:experiment}

After constructing our datasets, we frame our evaluation into four research questions, as outlined below. 
\begin{itemize}[leftmargin=*]

\item \textbf{RQ1:} Can the graph learning-based method help improve the data collection efficiency?
\item \textbf{RQ2:} How various and representative are the collected security commits? 
\item \textbf{RQ3:} What are the unique patterns of security commits in Python? 
\item \textbf{RQ4:} How do the wild commit samples help improve \gnn{} model for downstream security commit detection? 
\end{itemize}

\subsection{Dataset Construction (RQ1)}\label{results:efficiency}

After keyword filtering and graph-based identification with humans in the loop, we collect 1,258 security commits in total. Specifically, as shown in Table~\ref{tab: dataset}, there are 729, 400, and 129 security commits in the base, pilot, and augmented datasets, respectively. Also, 2,791 non-security commits are manually labeled during the collection process.


\begin{table}[h]
\vspace{-0.05in}
\centering
    \caption{The composition of \db{}.}
    \setlength{\tabcolsep}{1.7mm}{
    \begin{tabular}{c|p{1.0cm}<{\centering}|p{1.0cm}<{\centering}|p{1.4cm}<{\centering}|p{1.0cm}<{\centering}}
    \toprule
    {\diagbox{\bf Commit}{\bf Dataset}} & {\bf Base} & {\bf Pilot} & {\bf Augmented} & {\bf Total} \\
    \midrule
    {\bf Security} & {729} &  {400} &  {129} & {1258} \\
    \midrule
    {\bf Non-Security} & {2134} & {535} & {122} &{2791} \\
    \bottomrule
    \end{tabular}
    }
    \label{tab: dataset}
\vspace{-0.05in}
\end{table}

Table~\ref{tab:spr} lists the augmentation efficiency of random selection, keyword filtering, and \gnn{}.
Compared with identifying security commits from scratch, the keyword filtering mechanism improves the collecting efficiency by over 30 percentage points and \gnn{} improves the efficiency by 40 percentage points. 

\begin{table}[ht]
\vspace{-0.05in}
\centering
\caption{Efficiency of keyword filtering and \gnn{}.}
\setlength{\tabcolsep}{4mm}{
\begin{tabular}{c|c|c|c}
\toprule
\textbf{Method}      & \textbf{\# Candidates} & \textbf{\# Verified SC$^{*}$} & \textbf{Ratio} \\
 \midrule
{Random~\cite{wang2021patchdb}} & {-} & {-} & {6-10\%}    \\ 
\midrule
{Keywords} & {935} & {400} & {42.70\%} \\ 
\midrule
{\gnn{}} & {251} & {129} & {51.39\%} \\ 
\bottomrule

\end{tabular}
}
\begin{tablenotes}[flushleft]
    \footnotesize
    \item $^{*}$ SC = Security Commits.
\end{tablenotes}
\label{tab:spr}
\vspace{-0.15in}
\end{table}

\begin{table}[]
\centering
\caption{Top 5 repositories by number of security commits.}
\label{tab:repo}
\setlength{\tabcolsep}{5.4mm}{
\begin{tabular}{c|c|c}
\toprule
\textbf{Repository} & \textbf{\#SecurityCommits} & \textbf{\textbf{Proportion}} \\
\midrule
django      & 166  & 13.20\%   \\ \midrule
twisted     & 87   & 6.91\%   \\ \midrule
glance      & 54   & 4.29\%     \\ \midrule
pillow     & 41   & 3.26\%     \\ \midrule
numpy       & 39   & 3.10\%        \\ \midrule
\rowcolor{gray!10}\textbf{Total of Top 5}                   & \textbf{387}   &   \textbf{30.76\%} \\
\bottomrule
\end{tabular}
}
\vspace{-0.1in}
\end{table}

\subsection{Security Commits Categorization and Distribution (RQ2)}

\begin{figure}[h]
\vspace{-0.1in}
    \centering
    \includegraphics[width=0.75\linewidth]{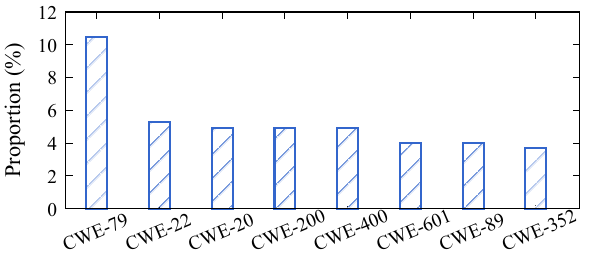}
    \vspace{-0.05in}
    \caption{The top 8 CWE types in \db{}.}
    \vspace{-0.1in}
    \label{fig:cwe}
\end{figure}

NVD CWE slice~\cite{CWE_slice} associated classification taxonomy serves to identify and describe security vulnerabilities.
To understand the purpose of these commits, we investigate the CWE types associated with the CVE reports and plot the distribution of the CWE types that have been explicitly documented. Among the 729 security commits linked to 556 CVEs, due to the limited number of MITRE human analysts, only 312 (56.1\%) CVEs have been assigned CWEs. 
Even so, there are already 119 distinct CWEs associated with our security commits in the base dataset, which means our \db{} contains at least 119 types of security commits in terms of corresponding vulnerabilities.
Figure~\ref{fig:cwe} enumerates the most common CWEs, including frequent security problems such as cross-site scripting (CWE-79), path traversal (CWE-22), etc. Note that we do not directly assign CWE type to security samples in the remaining base, pilot, and augmented dataset since the MITRE CWE team has its own internal process. However, based on our observation and our data collection approaches that are able to introduce wild security commits with more variance (as discussed in~\ref{exp:variance}), \db{} can encompass a broad range of security concerns with various kinds of security commits, including but not limited to above 119 CWEs. 


Our collected security commits distribute among 351 popular GitHub repositories unevenly. Among them, 69 repositories provide more than two security commits, bringing a certain amount of variety. In Table~\ref{tab:repo}, the top five repositories that have the most occurrence in our dataset are django~\cite{django_2023}, twisted~\cite{twisted_2023}, glance~\cite{openstack_2023}, pillow~\cite{python-pillow_2023}, and numpy~\cite{numpy_2023}, implying that the samples in \db{} align with the popularity trend of security issue in practice.



\subsection{Patch Patterns (RQ3)}
\label{rq3}

We manually go through the whole \db{} dataset, 
and discover four common security fix patterns (taking up 85.85\% of all security commit samples) that may benefit software maintenance, i.e., adding or updating sanity checks,  updating APIs, updating regular expressions, and updating security properties, as listed in Table~\ref{tab:pattern}.

\begin{table}[]
\centering
\caption{The pattern types of security commits in \db{}.}
\label{tab:pattern}
\setlength{\tabcolsep}{4.0mm}{
\begin{tabular}{l|c|c}
\toprule
    \textbf{Pattern} & \textbf{\#Commits} & \textbf{Proportion} \\ 
    \midrule
    {1) Add or Update Sanity Checks} & {416} & {37.12\%} \\ 
    \midrule
    {2) Update API Usage} & {241} & {19.16\%} \\ 
    \midrule
    {3) Update Regular Expressions} & {189} & {15.02\%} \\ 
    \midrule
    {4) Restrict Security Properties} & {183} & {14.55\%} \\ 
    \midrule
    {5) Others} & {178} & {14.15\%} \\ 
    \midrule 
    \rowcolor{gray!10}{\bf Total} & {\bf 1258} & {\bf 100\%} \\
    \bottomrule
\end{tabular}
}
\vspace{-0.1in}
\end{table}



\subsubsection{Add or Update Sanity Checks}
A sanity check is a basic method to quickly evaluate if a claim or a calculation result can be true, which has been extensively applied to multiple scenarios, e.g., authentication property verification, access control, HTTP request checking~\cite{wang2020machine}. 
We summarize three representative patterns that fix the vulnerabilities via adding or updating sanity checks, which are presented by 37.12\% of security commits in \db{}.

\noindent{\bf Authentication.} Authentication is the act of proving an assertion, e.g., we need to compare the identity with the system data to verify a system user. The authentication-related vulnerabilities
provide attackers the opportunities to masquerade as legitimate users. To defend them, an effective solution is to perform the additional authentication by adding more check requirements or making existing conditions more restrictive. List~\ref{lst:auth} presents an example of fixing an authentication vulnerability by narrowing down an existing restriction from \texttt{\small True} (i.e., all possible return values except \texttt{\small False}) to \texttt{\small "on"} only.

\lstdefinestyle{lst}{
    float=th,
    floatplacement=tbp,
    numbers=left, 
    numberstyle=\scriptsize, 
    numbersep = 5pt,
    framexleftmargin = 0in,
    framexrightmargin = 0in,
    breaklines = true,
    xleftmargin = 0.18in,
    xrightmargin = 0.1in,
    basicstyle=\ttfamily\scriptsize, 
    frame=lines,
    showtabs=true,
    showspaces=true,
    showstringspaces=false,
    literate={\ }{{\ }}1,
    aboveskip=-0.00in,
    belowskip=-0.15in,
}

\begin{lstlisting}[
language=diff, 
style=lst,
caption=An example of security commit to fix authentication vulnerability (CVE-2022-0273).,
label={lst:auth},
mathescape=true
]
 $\textbf{commit 0c0313f375bed7b035c8c0482bbb09599e16bfcf}$ 
 diff --git a/cps/shelf.py b/cps/shelf.py
 @@ -248,7 +248,7 @@ def create_edit_shelf(shelf,
 ...
         $\textbf{return}$ redirect(url_for('web.index'))
-    is_public = 1 if to_save.get("is_public") else 0
+    is_public = 1 if to_save.get("is_public") == "on" else 0
     $\textbf{if}$ config.config_kobo_sync:
 ...
\end{lstlisting}

\noindent{\bf Authorization.} Authorization refers to the process of granting or denying access to certain data or actions within a system.
Authorization comes after authentication and is achieved by an access control list (ACL).
The ACL is used to check the user identity with a list of authorized operations and determine which actions a user is allowed to take, e.g., file and data permission.
Unrestricted authorization may lead to improper resource consumption since attackers could bypass the system to access high-security level data. List~\ref{lst:access control1} is an example that fixes an authorization bypass exploit by requiring the value of \texttt{\small os.environ.get('GITHUB\_ACTIONS')} to be \texttt{\small true}.

\lstdefinestyle{lst}{
    float=th,
    floatplacement=tbp,
    numbers=left, 
    numberstyle=\scriptsize, 
    numbersep = 5pt,
    framexleftmargin = 0in,
    framexrightmargin = 0in,
    breaklines = true,
    xleftmargin = 0.18in,
    xrightmargin = 0.1in,
    basicstyle=\ttfamily\scriptsize, 
    frame=lines,
    showtabs=true,
    showspaces=true,
    showstringspaces=false,
    literate={\ }{{\ }}1,
    aboveskip=+0.10in,
    belowskip=-0.30in,
}

\begin{lstlisting}[
language=diff, 
style=lst,
caption=An example of security commit that fixes an authorization bypass exploit vulnerability (CVE-2022-46179).,
label={lst:access control1},
mathescape=true
]
 $\textbf{commit c658b4f3e57258acf5f6207a90c2f2169698ae22}$  
 diff --git a/core.py b/core.py
 @@ -112,7 +112,7 @@ def actualsys() :
     $\textbf{if}$ attemps == 6:
         ## Brute force protection
         $\textbf{raise}$ Exception("Too many password attempts.")
-    if os.environ.get('GITHUB_ACTIONS') != "":
+    if os.environ.get('GITHUB_ACTIONS') == "true":
         logging.warning("Running on Github Actions")
         actualsys()
     $\textbf{elif}$ uname == cred.name and pwdhash == cred.pass:
\end{lstlisting}

\noindent{\bf HTTP Request.} If the interpretation of Content-Length and/or Transfer-Encoding headers between HTTP servers are inconsistent, the attackers may take advantage of this issue and send malicious requests to the servers, i.e., HTTP request smuggling. 
A good solution is to maintain the same interpretation methods in both front-end and back-end servers. 
In this way, an effective coding practice is to add consistent sanity checks on request interpretation for both servers. 
List~\ref{lst:http} adds such a sanity check on \texttt{\small data} to determine if all characters are digits.

\lstdefinestyle{lst}{
    float=th,
    floatplacement=tbp,
    numbers=left, 
    numberstyle=\scriptsize, 
    numbersep = 5pt,
    framexleftmargin = 0in,
    framexrightmargin = 0in,
    breaklines = true,
    xleftmargin = 0.18in,
    xrightmargin = 0.1in,
    basicstyle=\ttfamily\scriptsize, 
    frame=lines,
    showtabs=true,
    showspaces=true,
    showstringspaces=false,
    literate={\ }{{\ }}1,
    aboveskip=-0.00in,
    belowskip=-0.15in,
}

\begin{lstlisting}[
language=diff, 
style=lst,
caption=An example of security commit that fixes an HTTP request smuggling vulnerability (CVE-2022-24801).,
label={lst:http},
mathescape=true
]
 $\textbf{commit 8ebfa8f6577431226e109ff98ba48f5152a2c416}$ 
 diff --git a/src/twisted/web/http.py b/src/twisted/web/http.py
 @@ -2274,6 +2274,8 @@ def fail():
     $\textbf{if}$ header == b"content-length":
+        if not data.isdigit():
+            return fail()
         $\textbf{try}$:
             length = int(data)
         $\textbf{except}$ ValueError:
\end{lstlisting}

\subsubsection{Update API Usage}
Compared with implementing the fixes from scratch, there are abundant well-formulated packages that can be adopted to realize the intended functionalities and help enforce security restrictions. 
We notice that a large number (19.16\%) of Python security commits fix vulnerabilities by imposing or substituting APIs. 
We further categorize such security fixes 
according to their application scenarios. 

\noindent{\bf General Purpose.} There is a set of security-related modifications on built-in packages shared by applications for various purposes. For instance, \texttt{\small re.escape} is an API to escape non-alphanumerics that are not part of regular expression syntax, to avoid OS command injection, code injection, and regular expression injection. List~\ref{lst:re} is a commit example to fix regular expression injection vulnerability, which demonstrates the application of \texttt{\small re.escape} on \texttt{\small user} and \texttt{\small collection\_url}.

\lstdefinestyle{lst}{
    float=th,
    floatplacement=tbp,
    numbers=left, 
    numberstyle=\scriptsize, 
    numbersep = 5pt,
    framexleftmargin = 0in,
    framexrightmargin = 0in,
    breaklines = true,
    xleftmargin = 0.18in,
    xrightmargin = 0.1in,
    basicstyle=\ttfamily\scriptsize, 
    frame=lines,
    showtabs=true,
    showspaces=true,
    showstringspaces=false,
    literate={\ }{{\ }}1,
    aboveskip=-0.00in,
    belowskip=-0.15in,
}

\begin{lstlisting}[
language=diff, 
style=lst,
caption=An example of security commit that fixes a regular expression injection vulnerability (CVE-2015-8748).,
label={lst:re},
mathescape=true
]
 $\textbf{commit 4bfe7c9f7991d534c8b9fbe153af9d341f925f98}$ 
 diff --git a/radicale/rights/regex.py b/radicale/rights/regex.py
 @@ -65,7 +65,10 @@ def _read_from_sections(user, collection_url, permission):
 ...
-    regex = ConfigParser({"login": user, "path": collection_url})
+    # Prevent "regex injection"
+    user_escaped = re.escape(user)
+    collection_url_escaped = re.escape(collection_url)
+    regex = ConfigParser({"login": user_escaped, "path": collection_url_escaped})
 ...
\end{lstlisting}

\noindent{\bf Web Applications.} To properly process the inputs of web applications, security commits can adopt 
APIs in third-party packages for Python
(e.g., \texttt{\small parser.quote}, \texttt{\small request.server.escape}, \texttt{\small django.utils.html.escape}, and \texttt{\small html.unescape}) to escape ampersands, brackets, and quotes to the HTML/XML entities or HTTP requests for defeating cross-site scripting (XSS) and HTTP Smuggling. 
List~\ref{lst:xss2} is an example that fixes an XSS vulnerability by using the API \texttt{\small django.utils.html.escape}.

\lstdefinestyle{lst}{
    float=th,
    floatplacement=tbp,
    numbers=left, 
    numberstyle=\scriptsize, 
    numbersep = 5pt,
    framexleftmargin = 0in,
    framexrightmargin = 0in,
    breaklines = true,
    xleftmargin = 0.18in,
    xrightmargin = 0.1in,
    basicstyle=\ttfamily\scriptsize, 
    frame=lines,
    showtabs=true,
    showspaces=true,
    showstringspaces=false,
    literate={\ }{{\ }}1,
    aboveskip=+0.00in,
    belowskip=-0.15in,
}

\begin{lstlisting}[
language=diff, 
style=lst,
caption=An example of security commit that fixes an XSS vulnerability (CVE-2022-24710).,
label={lst:xss2},
mathescape=true
]
 $\textbf{commit f6753a1a1c63fade6ad418fbda827c6750ab0bda }$
 diff --git a/weblate/trans/forms.py b/weblate/trans/forms.py
 @@ -37,6 +37,7 @@
 ...
+from django.utils.html import escape
 ...
-    label = str(unit.translation.language)
+    label = escape(unit.translation.language)
 ...
\end{lstlisting}

\noindent{\bf Shell Commands.} To handle the shell commands securely, security fixes can adopt \texttt{\small shlex.quote} and \texttt{\small subprocess} to load or execute the commands. 
With the \texttt{\small shlex.quote} API, we can have an escaped version of shell inputs, which can be safely used as tokens in a command line to avoid shell command injection.
List~\ref{lst:shell} is an example that shows the usage of \texttt{\small shlex.quote} to fix a shell injection vulnerability. 

\lstdefinestyle{lst}{
    float=th,
    floatplacement=tbp,
    numbers=left, 
    numberstyle=\scriptsize, 
    numbersep = 5pt,
    framexleftmargin = 0in,
    framexrightmargin = 0in,
    breaklines = true,
    xleftmargin = 0.18in,
    xrightmargin = 0.1in,
    basicstyle=\ttfamily\scriptsize, 
    frame=lines,
    showtabs=true,
    showspaces=true,
    showstringspaces=false,
    literate={\ }{{\ }}1,
    aboveskip=-0.00in,
    belowskip=-0.15in,
}

\begin{lstlisting}[
language=diff, 
style=lst,
caption=An example of security commit that fixes a shell injection vulnerability (CVE-2013-7416).,
label={lst:shell},
mathescape=true
]
 $\textbf{commit 2817869f98c54975f31e2dd674c1aefa70749cca }$
 diff --git a/canto_curses/guibase.py b/canto_curses/guibase.py
 @@ -156,6 +156,11 @@ def _fork(self, path, href, text, fetch=False):
 ...
+    href = shlex.quote(href)
 ...
\end{lstlisting}

\noindent{\bf Path Name.} 
If a path name is improperly neutralized, attackers may access the files and directories outside of the restricted location. 
This vulnerability can occur by using absolute file paths or manipulating the path variables where the reference files contain ``dot-dot-slash (../)" sequences or variations.
To effectively escape such unsafe sequences, Python security commits usually adopt the secure APIs, e.g., \texttt{\small werkzeug.utils.safe\_join}, \texttt{\small yaml.safe\_load}, and \texttt{\small werkzeug.utils.secure\_filename}, to prevent the files or directories from being accessed by malicious users. 
List~\ref{lst:path traversal} is a commit example that fixes a path traversal via using the API \texttt{\small werkzeug.utils.secure\_filename}.

\lstdefinestyle{lst}{
    float=th,
    floatplacement=tbp,
    numbers=left, 
    numberstyle=\scriptsize, 
    numbersep = 5pt,
    framexleftmargin = 0in,
    framexrightmargin = 0in,
    breaklines = true,
    xleftmargin = 0.18in,
    xrightmargin = 0.1in,
    basicstyle=\ttfamily\scriptsize, 
    frame=lines,
    showtabs=true,
    showspaces=true,
    showstringspaces=false,
    literate={\ }{{\ }}1,
    aboveskip=+0.10in,
    belowskip=-0.25in,
}

\begin{lstlisting}[
language=diff, 
style=lst,
caption=An example of security commit that fixes a path traversal vulnerability (CVE-2022-23609).,
label={lst:path traversal},
mathescape=true
]
 $\textbf{commit 1eb1e5428f0926b2829a0bbbb65b0d946e608593}$ 
 diff --git a/upload/server.py b/upload/server.py
 @@ -5,7 +5,7 @@
-
+import werkzeug.utils
 @@ -189,7 +189,7 @@ def uploadimage():
     filename = all_files[0][1] + all_files[0][2]
-    remove(filename)
+    remove(werkzeug.utils.secure_filename(filename))
     $\textbf{del}$ all_files[0]
     length = len(all_files)
\end{lstlisting}

\subsubsection{Update Regular Expressions}
Python has become a popular choice for back-end web development, and it is usually combined with some other front-end languages~\cite{python_app}. For this reason, we observe there are 15.02\% fixes that modify the regular expressions to avoid XSS, SQL injection, and open redirect vulnerabilities. 
The regular expression patterns are tailored to match specific strings within the given text, including SQL commands, URLs, and other scripts.

\noindent{\bf SQL Commands.} The improper neutralization of SQL commands may lead to SQL injection vulnerabilities, which allow attackers to manipulate the backend database and access the information not intended to be displayed.
The corresponding fixes need to escape the unsafe characters. 
List~\ref{lst:sql} is a fixed example of SQL injection vulnerability, which substitutes the matched single and double quote characters (i.e., \texttt{\small '} and \texttt{\small "}) in the string \texttt{\small self.queueid}.

\lstdefinestyle{lst}{
    float=th,
    floatplacement=tbp,
    numbers=left, 
    numberstyle=\scriptsize, 
    numbersep = 5pt,
    framexleftmargin = 0in,
    framexrightmargin = 0in,
    breaklines = true,
    xleftmargin = 0.18in,
    xrightmargin = 0.1in,
    basicstyle=\ttfamily\scriptsize, 
    frame=lines,
    showtabs=true,
    showspaces=true,
    showstringspaces=false,
    literate={\ }{{\ }}1,
    aboveskip=+0.0in,
    belowskip=-0.15in,
}

\begin{lstlisting}[
language=diff, 
style=lst,
caption=An example of security commit that fixes a SQL injection vulnerability (CVE-2014-125082).,
label={lst:sql},
mathescape=true
]
 $\textbf{commit fc2c1ea1b8d795094abb15ac73cab90830534e04}$
 diff --git a/.../model.py b/.../model.py
 @@ -772,13 +772,13 @@ def _get_filter(self):
 $\textbf{if}$ self.queueid:
-    ... = '%s'" % (self.queueid)
+    ... = '%s'" % (re.sub("[\"']", "", self.queueid))
\end{lstlisting}

\noindent{\bf URLs.} The improper neutralization of URLs may lead to open redirect vulnerability, which redirects an unsuspecting victim from a legitimate domain to an attacker’s phishing site. 
Effective mitigation is to replace the dangerous special characters with trusted symbols. List~\ref{lst:redirect} is an example of an open redirect vulnerability, which replaces the explicit backslash with an encoded backslash to circumvent the dangerous redirect.

\lstdefinestyle{lst}{
    float=th,
    floatplacement=tbp,
    numbers=left, 
    numberstyle=\scriptsize, 
    numbersep = 5pt,
    framexleftmargin = 0in,
    framexrightmargin = 0in,
    breaklines = true,
    xleftmargin = 0.18in,
    xrightmargin = 0.1in,
    basicstyle=\ttfamily\scriptsize, 
    frame=lines,
    showtabs=true,
    showspaces=true,
    showstringspaces=false,
    literate={\ }{{\ }}1,
    aboveskip=-0.00in,
    belowskip=-0.15in,
}

\begin{lstlisting}[
language=diff, 
style=lst,
caption=An example of security commit that fixes an open redirect vulnerability (CVE-2019-10255).,
label={lst:redirect},
mathescape=true
]
 $\textbf{commit 08c4c898182edbe97aadef1815cce50448f975cb}$ 
 diff --git a/auth/login.py b/auth/login.py
 @@ -39,6 +39,10 @@ def _redirect_safe(self, url, ...):
+    url = url.replace("\\", "%5C")
     parsed = urlparse(url)
     $\textbf{if}$ parsed.netloc $\textbf{or not}$ (parsed.path + '/').startswith(self.base_url):
\end{lstlisting}

\noindent{\bf Scripts.} The improper input validation and encoding during web page generation may lead to XSS, which is able to reveal the cookies, session tokens, or other sensitive information retained by the browser to the attackers. A straightforward solution is to validate the matched characters of a pre-defined pattern. List~\ref{lst:xss} is an example to fix the XSS vulnerability by re-matching the characters between parentheses instead of the characters between square brackets and validating the matched pattern one by one.

\lstdefinestyle{lst}{
    float=th,
    floatplacement=tbp,
    numbers=left, 
    numberstyle=\scriptsize, 
    numbersep = 5pt,
    framexleftmargin = 0in,
    framexrightmargin = 0in,
    breaklines = true,
    xleftmargin = 0.18in,
    xrightmargin = 0.1in,
    basicstyle=\ttfamily\scriptsize, 
    frame=lines,
    showtabs=true,
    showspaces=true,
    showstringspaces=false,
    literate={\ }{{\ }}1,
    aboveskip=+0.10in,
    belowskip=-0.25in,
}

\begin{lstlisting}[
language=diff, 
style=lst,
caption=An example of security commit that fixes an XSS vulnerability (CVE-2021-3994).,
label={lst:xss},
mathescape=true
]
 $\textbf{commit a22eb0673fe0b7784f99c6b5fd343b64a6700f06}$ 
 diff --git a/helpdesk/models.py b/helpdesk/models.py
 @@ -238 +238 @@ def cvesForCPE(cpe,
     $\textbf{if not}$ text:
         $\textbf{return}$ ""
-    pattern = fr'([\[\s\S\]]*?)\(([\s\S]*?):([\[\s\S\]]*?)\)'
+    pattern = fr'([\[\s\S\]]*?)\(([\s\S]*?):([\s\S]*?)\)'
     # Regex check
     $\textbf{if}$ re.match(pattern, text):
         # get get value of group regex
\end{lstlisting}

\subsubsection{Restrict Security Properties} 
The exploits often result from improper settings of security properties. 
14.55\% security commits in \db{} fix improper settings by updating boolean flags from \texttt{\small True} to \texttt{\small False} or vice versa, adding more arguments to methods, or adding security decorators.

\noindent{\bf Update Security Flags.} 
Security flags perform restrictions on the methods that may have access to sensitive objects. 
Improper restrictions on such flags may expose users to a risky environment and/or lead to sensitive information leakage. 
List~\ref{lst:flag} changes the flag from \texttt{\small False} to \texttt{\small True} to fix a vulnerability, where a sensitive cookie does not have a `HttpOnly' flag.

\lstdefinestyle{lst}{
    float=th,
    floatplacement=tbp,
    numbers=left, 
    numberstyle=\scriptsize, 
    numbersep = 5pt,
    framexleftmargin = 0in,
    framexrightmargin = 0in,
    breaklines = true,
    xleftmargin = 0.18in,
    xrightmargin = 0.1in,
    basicstyle=\ttfamily\scriptsize, 
    frame=lines,
    showtabs=true,
    showspaces=true,
    showstringspaces=false,
    literate={\ }{{\ }}1,
    aboveskip=-0.00in,
    belowskip=-0.15in,
}

\begin{lstlisting}[
language=diff, 
style=lst,
caption=An example of security commit that fixes a vulnerability where the sensitive cookie does not have a `HttpOnly' flag (CVE-2019-25091).,
label={lst:flag},
mathescape=true
]
 $\textbf{commit 60a3fe559c453bc36b0ec3e5dd39c1303640a59a}$ 
 diff --git a/src/nsupdate/settings/base.py b/src/nsupdate/settings/base.py
 @@ -283,7 +283,7 @@
 ...
-CSRF_COOKIE_HTTPONLY = False
+CSRF_COOKIE_HTTPONLY = True
 ...
\end{lstlisting}

\noindent{\bf Add Restriction Arguments.} Some restriction arguments will be passed to the functions during execution. Improper argument settings may lead to a variety of mishandling. As shown in List~\ref{lst:arg}, the \texttt{\small formaction} is added to restrict the attributes of a variable to avoid XSS vulnerability.

\lstdefinestyle{lst}{
    float=th,
    floatplacement=tbp,
    numbers=left, 
    numberstyle=\scriptsize, 
    numbersep = 5pt,
    framexleftmargin = 0in,
    framexrightmargin = 0in,
    breaklines = true,
    xleftmargin = 0.18in,
    xrightmargin = 0.1in,
    basicstyle=\ttfamily\scriptsize, 
    frame=lines,
    showtabs=true,
    showspaces=true,
    showstringspaces=false,
    literate={\ }{{\ }}1,
    aboveskip=-0.00in,
    belowskip=-0.15in,
}

\begin{lstlisting}[
language=diff, 
style=lst,
caption=An example of security commit that fixes a cross-site-scripting (XSS) vulnerability (CVE-2021-28957).,
label={lst:arg},
mathescape=true
]
 $\textbf{commit 10ec1b4e9f93713513a3264ed6158af22492f270}$ 
 diff --git a/src/lxml/html/defs.py b/src/lxml/html/defs.py
 @@ -23,6 +23,8 @@
 ...
+    # HTML5 formaction
+    'formaction'
     ])
 ...
\end{lstlisting}

\noindent{\bf Add Security Decorators.} A decorator is a function that takes another function and extends the behavior of the function without explicit modification. This mechanism has been widely adopted by security commits to add more detailed security restrictions on existing methods. List~\ref{lst:access control2} shows a security commit that fixes an access control vulnerability by adding decorator \texttt{\small security.private} to function \texttt{\small enumerateRoles}.

\lstdefinestyle{lst}{
    float=th,
    floatplacement=tbp,
    numbers=left, 
    numberstyle=\scriptsize, 
    numbersep = 5pt,
    framexleftmargin = 0in,
    framexrightmargin = 0in,
    breaklines = true,
    xleftmargin = 0.18in,
    xrightmargin = 0.1in,
    basicstyle=\ttfamily\scriptsize, 
    frame=lines,
    showtabs=true,
    showspaces=true,
    showstringspaces=false,
    literate={\ }{{\ }}1,
    aboveskip=+0.10in,
    belowskip=-0.25in,
}

\begin{lstlisting}[
language=diff, 
style=lst,
caption=An example of security commit that fixes an access control vulnerability (CVE-2021-21336).,
label={lst:access control2},
mathescape=true
]
 $\textbf{commit 2dad81128250cb2e5d950cddc9d3c0314a80b4bb}$ 
 diff --git a/src/Products/plugins/ZODBRoleManager.py b/src/Products/plugins/ZODBRoleManager.py
 @@ -112,6 +112,7 @@ def getRolesForPrincipal(self, principal, request=None):
     #   IRoleEnumerationPlugin implementation
+    @security.private
     $\textbf{def}$ enumerateRoles(self, id=None, exact_match=False, sort_by=None, max_results=None, **kw):
         """ See IRoleEnumerationPlugin.
\end{lstlisting}

\subsection{Unique Patterns Captured from the Wild (RQ4)}\label{exp:variance}

Recall that we construct pilot and augmented datasets because the base dataset provides a limited number of security commits samples. Here, we further show the examples captured by our security commit collection approaches that introduce more variety in syntax and semantics of security-related code changes, enabling wider applications of \db{} in solving real-world Python-related security issues.

\subsubsection{Data Variety Introduced by Pilot Dataset}
We study the contribution of involving the pilot dataset for \gnn{} by comparing the model trained only on the base dataset and the model trained on the combination of the base and pilot datasets.
We find that the pilot dataset helps the latter model to be able to identify more wild security commits. For instance, the latter \gnn{} can detect more subtle changes. 
In List~\ref{lst:pilot}, the \texttt{\small '\%s'} has been changed to \texttt{\small ?} in a SQL query, protecting the database from being injected. 
The capability of detecting such minor changes is enabled by similar samples in the pilot dataset but not existed in the base dataset.

\lstdefinestyle{lst}{
    float=th,
    floatplacement=tbp,
    numbers=left, 
    numberstyle=\scriptsize, 
    numbersep = 5pt,
    framexleftmargin = 0in,
    framexrightmargin = 0in,
    breaklines = true,
    xleftmargin = 0.18in,
    xrightmargin = 0.1in,
    basicstyle=\ttfamily\scriptsize, 
    frame=lines,
    showtabs=true,
    showspaces=true,
    showstringspaces=false,
    literate={\ }{{\ }}1,
    aboveskip=-0.00in,
    belowskip=-0.15in,
}

\begin{lstlisting}[
language=diff, 
style=lst,
caption=An example of security commit detected by \gnn{} trained on the base and pilot datasets.,
label={lst:pilot},
mathescape=true
]
 $\textbf{commit 9d8adbc07c384ba51c2583ce0819c9abb77dc648}$ 
 diff --git .../__init__.py .../__init__.py
 @@ -71,7 +71,7 @@ def klauen(self,
-    a = u"name == '%s' AND item =='%s'" % (name, item)
+    a = u"name == ? AND item ==?", (name, item)
\end{lstlisting}

\subsubsection{Variance Introduced by Augmented Dataset}
We further evaluate to show that our augmented dataset can help train a model that is able to identify more various security commits from the wild. For example, after introducing augmented dataset into the training phase, the model detects a new escape pattern. As shown in List~\ref{lst:augmented}, the characters \texttt{\small <}, \texttt{\small >}, and \texttt{\small \&} have been escaped by being translated into Unicode, which prevents cross-site-scripting crafted with a partial JSON-serializable object. Compared with the escape expressions in Section~\ref{rq3} that only include ASCII characters, the augmented dataset help \gnn{} generalize the escapes to Unicode.


\lstdefinestyle{lst}{
    float=th,
    floatplacement=tbp,
    numbers=left, 
    numberstyle=\scriptsize, 
    numbersep = 5pt,
    framexleftmargin = 0in,
    framexrightmargin = 0in,
    breaklines = true,
    xleftmargin = 0.18in,
    xrightmargin = 0.1in,
    basicstyle=\ttfamily\scriptsize, 
    frame=lines,
    showtabs=true,
    showspaces=true,
    showstringspaces=false,
    literate={\ }{{\ }}1,
    aboveskip=-0.00in,
    belowskip=-0.15in,
}

\begin{lstlisting}[
language=diff, 
style=lst,
caption=A security commit example detected by the \gnn{} trained on the base{,} pilot{,} and augmented datasets.,
label={lst:augmented},
mathescape=true
]
 $\textbf{commit d3e428a6f7bc4c04d100b06e663c071fdc1717d9}$ 
 diff --git a/.../djblets_js.py b/.../djblets_js.py 
 @@ -28,11 +28,18 @@
+_safe_js_escapes = {
+    ord('&'): u'\\u0026',
+    ord('<'): u'\\u003C',
+    ord('>'): u'\\u003E',
+}
\end{lstlisting}
\vspace{-0.03in}
\section{Discussion}
\label{sec:discussion}

\noindent{\bf Usability.} The \gnn{} is versatile and not tied to any specific platform or commit format, making it compatible with a wide range of version control systems like GitHub and GitLab. By integrating the \gnn{} as an extension, contributors can easily add vulnerability fix labels to their commits, streamlining the code auditing process and reducing the workload on developers. In addition, downstream developers and users who use third-party libraries can benefit from the \gnn{} by being reminded to make necessary security fixes on time. 
Finally, researchers can obtain labeled commits without requiring extensive manual labor for future data-driven vulnerability and patch-related research.


\noindent{\bf Ethical Consideration.} The \gnn{} has the potential to identify undisclosed vulnerability fixes, but this presents a mixed blessing as attackers could exploit this information to target unpatched systems. Our objective in this paper is to prioritize the security of the users' systems; that is why we only share detailed information on the security fixes, rather than the vulnerabilities. By taking this approach, attackers cannot leverage the \gnn{} to gain additional details on the vulnerabilities. However, with the \gnn{}, open-source software maintainers can quickly reveal vulnerabilities as soon as security fixes become public, improving the overall security of their software systems.






\vspace{-0.03in}
\section{Threats to Validity}
\label{sec:threats}


\noindent{\bf Threats to internal validity.} 
Internal bias and errors pose a significant risk in the dataset construction phase. 
The most essential threat is the exclusion of critical keywords or tokens that are important for identifying security-related commits. 
To address this issue, we propose an automated approach for learning security-related keywords. 
However, the current approach of topic models prioritizes the occurrence probabilities of words, which may lead to ignoring infrequent but essential words or tokens. Additionally, commits lacking proper documentation or commit messages are often overlooked, leading to further bias in the dataset.




\noindent{\bf Threats to external validity.} 
Our experiment and dataset are focused on Python commits, which may limit the generalizability of the \gnn{} to other programming languages. Also, since our dataset derives from open-source software, the data may not be applicable for identifying security-related commits in closed-source systems. However, we aim to expand the scope of our research in the future by incorporating more programming languages and diversifying our dataset to enhance the applicability of the \gnn{}.


\noindent{\bf Threats to construct validity.} 
\gnn{} is built on top of Joern~\cite{joern} to construct the code property graphs for the programs of previous and current versions; thus, \gnn{} inherits the limitations of Joern. Since Joern disregards the calling relations among multiple functions, \gnn{} cannot handle the commits that only change the function calls. Besides, Joern cannot identify the import and use operations of Python packages; thus, \gnn{} discards these edges in \cpg{} when the commits only operate packages.



\vspace{-0.03in}
\section{Conclusion and Future Work}
\label{sec:conclusion}
By fully leveraging the commit message and the code change semantics, we construct a large-scale Python security commit dataset named \db{} that consists of three parts: base, pilot, and augmented datasets. 
To enrich the base dataset extracted from CVE reports, the pilot dataset collects the commits that contain the pre-defined security keywords in their commit messages. 
Given the diversified data samples, we further train \gnn{} to learn the security semantics in the code changes to compensate for the poorly documented commits. 
We conduct a large-scale empirical study of security commits by analyzing \db{} of 119 CWE categories across 351 repositories. 
The summarized patterns can assist further software maintenance, e.g., auto program repair.
In the future, we will adopt large language models to expand the dataset and apply the summarized patterns to fix the vulnerabilities automatically.

\section*{Acknowledgments}
We would like to thank our anonymous reviewers for their valuable comments and suggestions. This work was partially supported by the US Office of Naval Research grants N00014-23-1-2122.

\bibliographystyle{ieeetr}
\bibliography{reference}

\end{document}